\documentclass[11pt]{article}
\usepackage{amsmath,amssymb, mcite}
\usepackage{graphicx}
\usepackage{multirow}

\def\beq{\begin{equation}}
\def\eeq{\end{equation}}
\def\bea{\begin{eqnarray}}
\def\eea{\end{eqnarray}}
\def\ba{\begin{array}}
\def\ea{\end{array}}
\def\bit{\begin{itemize}}
\def\eit{\end{itemize}}
\def\pa{\partial}
\def\na{\nabla}
\def\nn{\nonumber}

\def\deo{\mathring{\delta}}

\def\al{\alpha}
\def\ga{\gamma}
\def\Ga{\Gamma}
\def\si{\sigma}

\def\de{\delta}

\def\b{\beta}

\def\la{\lambda}

\def\vp{\varphi}

\def\na{\nabla}
\def\rd{{\rm d}}

\def\lPL{\ell_{\rm Pl}}
\def\ePL{\eta_{\rm Pl}}

\def\b{\beta}

\def\vp{\varphi}

\def\bx{{\bf{x}}}
\def\by{{\bf{y}}}

\def\Seff{S_{\rm anom}}
\def\Feff{{\Gamma}_{\rm anom}}
\def\rd{{\rm d}}

\def\cO{{\cal O}}

\def\trad{t_{\rm rad}}
\def\etarad{\eta_{\rm rad}}

\def\Pan{\bigtriangleup_{\rm P}}
\def\GPan{{\rm G}_{\rm P}}

\def\GB{{{\rm E}_4}}

\begin{document}
\begin{center}
{\bf \Large Conformal anomalies and the Einstein Field Equations}\\[1.5cm]

{\bf Hadi Godazgar$^1$, Krzysztof A. Meissner$^2$ and Hermann Nicolai$^1$}

\vspace{1cm}
{{\it 
$^1$ Max-Planck-Institut f\"ur Gravitationsphysik
(Albert-Einstein-Institut)\\
M\"uhlenberg 1, D-14476 Potsdam, Germany\\[1mm]
$^2$ Faculty of Physics,
University of Warsaw\\
Pasteura 5, 02-093 Warsaw, Poland\\
}}
\end{center}

\vspace{1cm}

{\footnotesize
We compute corrections to the Einstein field equations which are induced by the
anomalous effective actions associated to the type $A$ conformal anomaly, both for 
the (non-local) Riegert action, as well as for the local action with dilaton. In all cases
considered we find that these corrections can be very large.}

\vspace{1cm}

\section{Introduction}
In this paper we study the corrections to Einstein's equations that originate from 
`anomalous' (local and non-local) effective actions whose variation gives rise to 
conformal anomalies, and investigate whether such corrections can lead to observable
corrections of the Einstein field equations. In previous work \cite{MN} it was argued that 
the resulting corrections can, in fact, be rather large  {\em if}  the cumulated 
effects of non-localities can overwhelm the smallness of the Planck scale $\lPL$ 
that normally suppresses higher order effects in (quantum) gravity. 
Here we will present more detailed calculations in support of this assertion, considering both 
an explicit closed form non-local action $\Feff[g]$ \cite{Riegert} (the `Riegert action'), as well 
as a local version $\Seff[g,\tau]$ with a dilaton field $\tau$; the latter is relevant for theories with 
spontaneously broken conformal symmetry \cite{ST,KS}.

To begin, let us briefly recall some basic properties of the conformal anomaly, see
refs. \cite{Deser2,Duff1,CD,FT,DS,EO,Deser1, Duffrev} for the original papers and further details. 
Generally speaking, conformal anomalies have two sources, namely the fact that the 
UV regulator for any conformal matter system coupled to gravity necessarily breaks 
conformal invariance, and secondly the fact that even for classically conformal theories 
the functional measure depends on the metric in a non-local manner. In four dimensions 
the anomaly takes the form 
\beq
{\cal A}=T^\mu_{\; \; \mu}=\frac{1}{180(4\pi)^2}\left(c \, C^2+a \, \GB \right)
\label{anom}
\eeq\\[-4mm]
where 
\bea\label{CGB}
C^2&\equiv&C_{\mu\nu\rho\si}C^{\mu\nu\rho\si}=R_{\mu\nu\rho\si}
R^{\mu\nu\rho\si}-2R_{\mu\nu}R^{\mu\nu}+\frac13 R^2\nn\\
\GB &=& R_{\mu\nu\rho\si}
R^{\mu\nu\rho\si}-4R_{\mu\nu}R^{\mu\nu}+R^2
\eea
and the coefficients $c$ and $a$ depend on the type of matter that is (conformally) coupled 
to gravity. $\GB$ is the Gauss-Bonnet density, a total derivative that gives a topological invariant
when integrated over a 4-dimensional manifold. A further possible contribution proportional to 
$\Box R$ to (\ref{anom}) can be dropped as it is obtainable by variation of a local 
functional. In the remainder of this paper we will be mainly concerned with the
Gauss-Bonnet invariant $\GB$, the type $A$ anomaly.

Although the anomaly (\ref{anom}) is a local expression, it is well known 
that it cannot be obtained by variation of a local action functional in terms of the
metric alone; however, in the case of spontaneously broken conformal symmetry
there is also a local version if there is an extra field, the associated Goldstone 
boson, {\it alias} the dilaton $\tau$. In the former case
we are thus concerned with the {\em anomalous} part $\Feff[g]$ of the 
non-local effective action that gives rise to (\ref{anom})  via variation of the 
conformal factor,
\beq\label{Fanom}
{\cal A} (x) = 
- \frac2{\sqrt{-g(x) }} g_{\mu\nu}(x) \frac{\delta \Feff [g]}{\delta g_{\mu\nu} (x)} 
\eeq
In general, $\Feff[g]$ will  be a rather complicated functional, involving 
an infinite series of products of curvature tensors interspersed with Green's
functions that themselves depend on the metric. Various candidate actions have 
been proposed and discussed in the literature. These are either of non-local type
\cite{Riegert,DS,BGVZ,Barvinsky:1995it, Mirzabekian:1995qf}, or, 
for spontaneously broken conformal symmetry, of local type with a 
dilaton \cite{Mott,ST,KS} (see also \cite{BV} for an introduction to covariant perturbation
theory and heat kernel techniques); however, even for local actions
the non-localities re-appear after elimination of the dilaton by its equations of motion.
For $\GB$ there is a closed form action 
(`Riegert action') analogous to the Polyakov action involving the inverse of the 
4th order Paneitz differential operator \cite{FradkinTseytlin, Paneitz, Riegert} which however does 
not produce conformally covariant correlators in the flat space limit \cite{OP,EO,Erdmenger} 
(in fact, no closed form action with this property seems to be known). In general,
$\Feff$ will also excite new (longitudinal) gravitational degrees of freedom, hence
yield a variant of `modified gravity'.

The anomalous action $\Feff$ complements the usual effective action, which is 
also non-local, but it is important to keep in mind the following points. First of all
the separation of the effective action into an anomalous and a non-anomalous
part is, of course, ambiguous as we are free to add and subtract any
Weyl invariant functional. Furthermore, if the anomalous action $\Feff$ is to be
determined only by `solving' the functional differential equation (\ref{Fanom}) 
it is clear that this equation is underdetermined, as it is only one condition
on a functional that depends on ten metric components.
Secondly, we should distinguish between the quantum effective action, which is
the generating functional of the one-particle irreducible Green's functions, and the 
Wilsonian effective action that is obtained by integrating out only those modes whose 
mass scale lies well above the energy scales one is considering. The latter can be 
considered as providing $\cO(\hbar)$ corrections to the classical field equations 
which are generally very small (such as for instance the modifications to Maxwell's 
equations induced by the Euler-Heisenberg Lagrangian) because the UV degrees of 
freedom are supposed to decouple from the IR degrees of freedom in this limit. The 
only exception to this rule concerns anomalies, which can be viewed as IR manifestations 
of trans-Planckian physics (which is otherwise `invisible' in low energy processes). One well known 
example is the origin of axion-gluon and axion-photon couplings which are generally 
thought to arise from integration over super-heavy quarks (after spontaneous
breaking of chiral symmetry) but which survive without suppression to the 
lowest energy scales. 

The possible physical consequences of conformal anomalies have been 
analyzed in great detail by E.~Mottola who has repeatedly emphasized the need 
to include the non-local contribution from $\Feff$ on the right-hand side of 
Einstein's equations (see (\ref{MEFT}) below) in an effective field theory 
approach \cite{Mott,MottReview,Mottola06,Giannotti08, Mott1}. Most recently, 
ref.~\cite{Mott1} studies the effect of the Riegert action in a local version of $\Feff$
with a `conformalon' field (to be distinguished from the dilaton, the Goldstone
boson of spontaneously broken conformal symmetry). In the weak field limit
there arise  modifications of gravitational waves by scalar modes arising from the 
mixing of the conformal metric degree of freedom with the conformalon; being based 
on an analysis of the corrected Einstein equations in a weak field approximation 
\cite{Mott1}  reaches the conclusion that the effects resulting from the conformal anomaly 
remain tiny (and thus compatible with present observations). However, as we will show,
in the presence of persistent sources it turns out that there can also be very large 
effects caused by integrating the non-localities back in time. More specifically we find 
that the terms quadratic in the Riemann tensor that were negligible in the analysis 
of ref.~\cite{Mott1} give the main contribution (as opposed to the $\Box R$ contribution 
which played the dominant role there), in conjunction  with the huge factor resulting 
from integrating `over a long time'. Because the corrections need not remain 
small, as we will show in explicit examples, they can potentially invalidate any 
given solution of Einstein's equations.  It is crucial here that our whole analysis 
takes place in a regime throughout which gravity can be treated as fully classical, and 
which remains well within the presumed range of validity of effective field 
theory.\footnote{Adopting a very different perspective, one might instead add 
the non-local action $\Feff$ to the {\em quantum} action, thereby cancelling
the anomaly, so as to end up with an exactly conformal theory at the quantum
level -- this would be the analog of the Polyakov program leading to Liouville
theory \cite{P}. However, unlike for $D=2$ where the action for the conformal 
degree of freedom is quadratic, one might anticipate problems with ghosts and 
unitarity in $D=4$ as the action for the conformalon is of 4th order in derivatives.}

Readers might wonder why no such effects are seen (or even discussed) for the 
anomalies known from particle physics. Since gauge anomalies cancel in the 
standard model, there is evidently no need to discuss for them possible non-local effects 
of the type considered here. However, this is less obvious for global anomalies, 
where, for example, one might ask about possible modifications to the Maxwell equations
resulting from the axial anomaly. The main difference is that in electrodynamics
these effects remain small, as in electrodynamics there is no analog  of the 
enormous increase in curvature as one moves back in time (in fact, the electromagnetic
fields in the universe are vanishingly small throughout its history). Therefore, while providing important corrections such as in the well-known decay of $\pi^0$, the anomaly-induced terms do not overwhelm the action.

\section{Anomaly induced modifications: why they can be large}

To investigate the effect of conformal anomalies on the gravitational field equations
we observe that, with a non-local action $\Feff$ depending only on $g_{\mu\nu}$,
these take the form
\beq\label{MEFT}
\lPL^{-2} \left(R_{\mu\nu} -\frac12 g_{\mu\nu} R \right)(x) =  
\,- \, \frac2{\sqrt{-g}} \frac{\delta \Feff [g]}{\delta g^{\mu\nu}(x)}  + \cdots 
\eeq
where the dots stand for matter contributions as well as contributions 
from non-anomalous higher order curvature corrections (we will use the
labels $\lPL$ and $\ePL$ for the Planck length and Planck time interchangably).
This equation entails the consistency condition 
\beq\label{DivF}
\na^\mu \left(
\, \frac2{\sqrt{-g(x)}} \frac{\delta \Feff [g]}{\delta g^{\mu\nu}(x)} \right) \,=\, 0
\eeq
which is automatically satisfied also for non-local $\Feff[g]$ if it is diffeomeorphism 
invariant. The vanishing divergence of the contribution from the non-anomalous part of 
the effective action is guaranteed by the standard Ward-Takahashi identities. 
Our main interest is thus in the variation of $\Feff$ w.r.t. metric deformations which 
are  neither trace nor diffeomorphisms; we will generically refer to such 
deformations as `gravitational waves' -- in the broadest possible  sense that these 
must satisfy the conditions at infinity first identified by Trautman \cite{Tr}.  
We emphasize that among the omitted terms (indicated by ellipses) on 
the right-hand side of  (\ref{MEFT}) all terms other than matter sources (in particular 
local higher order curvature corrections) are completely negligible at the present epoch.  
The crucial question is then whether the extra corrections induced by $\Feff$ remain 
small under all conceivable circumstances.

Given that these corrections must be taken into account there are two possibilities.
The first is that the corrections can indeed be shown to be very small for any type of application
(cosmological solutions, primordial fluctuations, black holes, gravitational waves, {\it etc.}). 
In this case one can reasonably resort to a weak field approximation. 
The consistency of any given solution of Einstein's equations can be ascertained 
by treating $\Feff$ as a perturbation, substituting the relevant solution into the right-hand side of 
(\ref{MEFT}) and checking whether the corrections indeed remain small; this is what we 
will do below. Otherwise we are faced 
with the problem of solving a hopelessly complicated partial integrodifferential equation
where the metric that is to be solved for not only appears via higher powers of the 
Riemann tensor but also in the Green's functions. A further complication is that
in this case we would not even know how to properly set up an initial 
problem.\footnote{As a technical aside, we note that for the corrections we will always 
employ the retarded Green's functions, as this is the only choice which is compatible 
with causality, hence physically reasonable, see also remarks at the beginning of 
section~3.1 below. The consistency of this prescription is also evident from the
fact that no causality problems arise with the local actions.}

As a `testbed' for our analysis we will employ conformally flat metrics 
\beq\label{ds}
 ds^2 = a^2(\eta) \big( -d\eta^2 + d\bx^2\big)
\eeq
which play a central role in cosmological applications of Einstein's equations.
Although this choice of line element greatly simplifies the analysis, we do
not expect that the incorporation of spatial inhomogeneities will alter our conclusions
in any significant way. One main advantage of the simple form of (\ref{ds}) 
is that the Green's functions needed for our 
computations are known explicitly for {\em any} profile of the scale factor $a(\eta)$,
hence exact computations are possible. In a conformally flat background, we can also ignore the the Weyl-squared anomaly---so we will only consider the type $A$ anomaly in our analysis. For the quadratic conformal d'Alembertian we have
\beq\label{Green1}
\sqrt{-g} \left( - \Box + \frac16 R(x)\right) G(x) = \de^{(4)} (x)
\eeq
where $\Box\equiv \Box_g$ is always understood to be the operator associated to the metric 
$g_{\mu\nu}(x)$. The retarded Green's function for the metric (\ref{ds}) is \cite{Waylen}
\beq\label{G2}
G\big(\eta,\bx;\eta',\by \big) = \frac{1}{4\pi |\bx - \by|} \cdot 
   \frac{\delta(\eta - \eta' - |\bx - \by|)}{a(\eta) a(\eta')}
\eeq
Secondly, we have the 4th order Paneitz operator \cite{Paneitz}
\beq\label{Pan}
\Pan = \sqrt{-g} \left( \Box \Box + 
2 \na_\mu \left[ \left( R^{\mu\nu} - \frac13 g^{\mu\nu} R\right) \na_\nu\right]  \right)
\eeq
which is also conformal (cf. section~2). The associated Green's function is defined by
\beq\label{Pan1}
\Pan \GPan(x) = \delta^{(4)}(x).
\eeq
Its explicit retarded version for (\ref{ds}) is even simpler than (\ref{G2}), {\it viz.}
\beq
G_{\rm P}(\eta,\bx;\eta',\by) = \frac1{8\pi} \theta \big(\eta - \eta' - |\bx - \by|\big)
\eeq
where $\theta$ is the step function ($+1$ for positive values of the argument, and 
zero otherwise). In particular, this Green's function is completely independent of the 
scale factor $a(\eta)$. 

To gain some quantitative insight and in order to study how the anomalous effective action 
affects cosmological solutions we will further specialise to the physically relevant case 
of the radiation era, where   
\begin{equation}\label{aeta}
 a(\eta)  = \frac{\eta}{\etarad} = \left(\frac{t}{\trad}\right)^{1/2}
\end{equation}
and the conformal time $\eta$ is related to cosmological time $t$ by 
$\eta = 2\sqrt{t \trad}$ where $\eta = \etarad = 2\trad$ corresponds to the 
end of the radiation era  ($\trad \sim 380\,000$ years).
The enhancement effect will then result from integrating back in time to $\eta=L$, which
we will `measure' in units of the conformal Planck time $\ePL = 2\sqrt{t_{\rm Pl} \trad}$
\beq\label{L}
L = n_* \ePL
\eeq
Because $\etarad/\ePL \sim 10^{28}$ we will see that we can take $n_*$ rather large ({\it e.g.}
$n_* \sim 10^{7}$ for the exit from inflation) and still pick up a very large contribution
for the ratio $\etarad/L$.
The enhancement is thus due to both the increase in curvature (the square of the
Riemann tensor behaves like $\sim \eta^{-8}$) towards $\eta = L$ {\em and} the 
huge factor $\etarad/L$; a further enhancement results from the inverse factor
$a(\eta')^{-1}$ in (\ref{G2}) (where $\eta'$ is to be integrated over). We repeat that 
we do not even need to get close to the Planck regime to see this effect!

Let us also stress that the enhancement effects exhibited below cannot be simply 
dismissed by invoking unknown quantum gravity effects in a regime where the 
effective field theory approximation breaks down. First of all, our analysis stays 
sufficiently away from the Planck regime. Secondly, let us nevertheless assume that 
the large effect `now' were cancelled by an unknown contribution coming from the integration 
over the Planck regime. Then, as we go back in time this contribution to the integral stays 
the same whereas the contribution from the integrals contributing to (\ref{MEFT}) decreases 
to zero as $\eta\rightarrow L$, so that at time $\eta = L$ we would be left with the quantum
gravity induced term, and we would again have a large effect (presumably invalidating
much of standard cosmology). Therefore, we are assuming that  whatever the 
theory governing the Planck regime is it must lead to the observable almost homogeneous, 
isotropic cosmology whence we start our analysis.

\section{Riegert Action} \label{sec:riegert}

In two dimensions the conformal anomaly, which is the Euler term in two dimensions, 
the Ricci scalar, can be obtained by varying the Polyakov action \cite{P}
\begin{equation} \label{Polaction}
 \int d^2 x \int d^2 y \sqrt{-g(x)} \sqrt{-g(y)} R(x) G^{\Box}(x,y) R(y),
\end{equation}
where $G^{\Box}$ is the Green function for the conformally invariant $\Box$ operator in two dimensions. The variation of the Ricci scalar by Weyl rescalings parametrised by $\sigma$, in two dimensions, is given by $\Box \sigma,$ hence it is clear that varying the Polyakov action above by Weyl rescalings reproduces the anomaly. Furthermore, the associated correlation functions of the action have the correct conformal behaviour in the flat space limit~\cite{EO, Erdmenger}.

The structure of the Polyakov action can be generalised to higher dimensions. In particular, in four dimensions the Riegert action \cite{Riegert} manifests the structure of the Polyakov action:
\begin{equation} \label{Riegert}
 \int d^4 x \int d^4 y\,%\sqrt{-g(x)} \sqrt{-g(y)}  
 \mathcal{G}(x) \GPan (x,y) \mathcal{G}(y),
\end{equation}
where $\mathcal{G}$ is given by a modification of the Euler density,
\begin{equation} \label{mathcalGdef}
 \mathcal{G} = \sqrt{-g} \left(E_4 - \frac23 \Box R\right),
 \end{equation}
and  $\GPan$ is the Green function of the conformally invariant, fourth order Paneitz operator, 
equation \eqref{Pan1} \cite{FradkinTseytlin, Paneitz, Riegert}. The density $\mathcal{G}$ is defined so that its variation under Weyl rescalings is of the desired form, 
 \begin{equation} \label{varmcG}
   \de_{\sigma} \mathcal{G} = 4 \Pan \sigma.
 \end{equation}
 Therefore, given that $\GPan$ is the Green function of a conformally invariant operator 
 the variation of the Riegert action with respect to the conformal factor gives the integral over the density $\mathcal{G}$ -- the type $A$ anomaly plus $\Box R$, which can be removed by introducing a local action, \emph{viz.}
 \begin{equation}
 \int d^4 x \sqrt{-g} R^2.
 \end{equation}
 
The Riegert action, therefore, is precisely what is required for obtaining the anomaly. However, it does not have the correct analyticity structure \cite{Deser, EO, Erdmenger}, which means, in particular, 
that the associated correlation functions (of energy momentum tensor operators) are not conformally
covariant \cite{OP} in the flat space limit \cite{EO, Erdmenger}. Hence the Riegert action 
must be modified by Weyl-invariant terms that correct its analytical structure. When conformal symmetry is spontaneously broken the Riegert action still does not have the correct analytic 
structure~\cite{ST} -- in this case the correct action is known~\cite{ST} and we study it in the next section, \ref{sec:local}.  Let us also note that the Riegert action can be presented in a local form,
\beq\label{conformalon}
\int d^4 x \left( - 2  \phi \Pan \phi \, + \, \phi \, \mathcal{G} \right),
\eeq
by means of the `conformalon' $\phi$ (which transforms as $\delta\phi = \si$);
this is the form of the action studied in ref.~\cite{Mott1}.
That this action does not have the right analytic properties is already suggested by the
appearance of the fourth order kinetic operator in (\ref{conformalon}) which is expected to 
lead to the usual problems in the quantized theory (ghosts, {\it etc.}).

Despite the Riegert action not having the required analyticity structure, it is a closed form action that produces the anomaly. Therefore, it is possible to precisely determine whether the effects observed in \cite{MN} are also present here. We will also be able to determine what the leading terms are, which is relevant for the spontaneously broken case. 

\subsection{Variation of Riegert action}

Already before varying the action with respect to the metric, we observe that the action \eqref{Riegert} is such that the arguments of the Green function are symmetrised. Therefore, regardless of the causal properties of the Green function that we start with, the equations of motion will have both retarded and advanced contributions from the Green function, which renders the theory acausal -- what happens at a given moment in time depends on what will happen to the future of it. This is a generic feature of non-local actions. The resolution is that the action that we have is the generating functional for $|in\rangle$ to $|out\rangle$ scattering matrix elements, but these, not being physical observables, need not satisfy causality properties \cite{Barvinsky}. The expectation values of operator, on the other hand, must satisfy causality, but for this we require the in-in or Schwinger-Keldysh formalism whereby the contour is extended to include evolution back in time from $t=\infty$ to the original state. Since time-ordering is done on the extended contour, the action is supplemented by new terms in order to differentiate between fields on the forward and backward part of the contour---the propagator is now also a matrix because the time-ordering, depending on which part of the contour the fields are, is different~\footnote{For example, in the time-ordered product of two fields with one of the fields on the forward part of the contour and the other on the backward part of the contour, the latter field always comes after the former, independently of what time either is evaluated at.}. The upshot of this procedure is that once we have varied the action we replace all the Green functions by the ones with retarded boundary conditions~\cite{SoussaWoodard}. In this way causality is restored, and we will follow this procedure in finding the equations of motion from the Riegert action.

The variation of the Riegert action, \eqref{Riegert}, up to the subtlety discussed in the previous paragraph, is
\begin{align}
& \int d^4 x \int  d^4 y \left[ 2 \,  \de \mathcal{G}(x) \, \GPan(x,y) \,  \mathcal{G}(y) - \int d^4 z \, \mathcal{G}(x) \,  \GPan(z,x) \, ( \de\Pan)(z) \,  \GPan(z,y) \,  \mathcal{G}(y) \right],
 \label{varRiegert}
\end{align}
where $\GPan$ is the retarded Green function for the Paneitz operator defined by equation \eqref{Pan1} and $\de$ denotes variation with respect to the metric. 

We use the results summarised in appendix \ref{app:variations} and evaluate the variations in equation \eqref{varRiegert}.
The variation of $\mathcal{G}$, \eqref{mathcalGdef}, is given by
\begin{align}
  \frac{\de \mathcal{G}}{\sqrt{-g}}  &= - \frac23 \Box \left[ \left( g_{\al \b} \Box - \na_{\al} \na_{\b} \right) \de g^{\al \b} \right] - \na^{\mu} \na^{\nu} \Big[ \de g^{\al \b}\big( 4 R_{\mu \al\b \nu} + 4 R_{\mu \nu} g_{\al\b} - 8 R_{\mu \al} g_{\nu \b}  \notag\\[2mm]
  & \quad+ \frac{14}{3} R_{\al \b} g_{\mu \nu} - 2 R g_{\al \b} g_{\mu \nu}  + 2 R g_{\al \mu} g_{\b \nu} \big)  \Big] + \frac13 \na_{\mu} \bigg[ \Big( \na^{\mu} R g_{\al \b} - 2 \de^{\mu}_{\al} \na_{\b} R \Big) \de g^{\al \b} \bigg], \label{eqn:varcG2}
 \end{align}
which is a total derivative, as it should be  (this is a useful check on our computation).
For the variation of the Paneitz operator, \eqref{Pan}, we get
\begin{align}
  \frac{\de \Pan}{\sqrt{-g}}   &=  - \frac13 \na_{\al} \na_{\b} \na_{\mu} \left[\deo g^{\al \b} \na^{\mu} \right] + \Box \na_{\al} \left[ \deo g^{\al \b} \na_{\b} \right] + \na_{\al} \na_{\b} \left[ \deo g^{\al \b} \Box \right] + \Box \left[ \deo g^{\al \b} \na_{\al} \na_{\b} \right] \notag\\[3mm]
   &\quad - \frac43 \na_{\al} \na_{\mu} \left[ \deo g^{\al \b} \na^{\mu} \na_{\b}\right] + \na_{\al} \left[ \deo g^{\al \b} \left( 2 R_{\mu \b} \na^{\mu} + \na_{\b} \Box - \frac23 R \na_{\b} \right) \right] \notag \\[3mm]
 & \quad + \frac13 \na^{\mu} \left[ \deo g^{\al \b} \left( 4 R_{\mu \al \nu\b} \na^{\nu} - \na_{\mu} \na_{\al} \na_{\b} + 6 R_{\mu \al} \na_{\b} - 2 R_{\al \b} \na_{\mu} \right)\right], \label{varPan2}
\end{align}
where only the traceless part of the metric deviation appears in the variation of $\Pan$:
\begin{equation}
 \deo g^{\al \b} = \de g^{\alpha \beta} - \frac14 g^{\alpha \beta} g_{\gamma \de} \de g^{\gamma \de}.
\end{equation}
The expressions are consistent with the variation of $\mathcal{G}$ and $\Pan$ with respect to the conformal factor. Namely, letting $ \de g_{\al \b} = 2 \sigma g_{\al \b}$, it is simple to verify that we recover equation \eqref{varmcG}. Moreover, since only the traceless variation of the metric appears in equation \eqref{varPan2}, the variation of the Paneitz operator with respect to the conformal factor vanishes -- consistent with the fact that it is a conformally invariant operator.

\subsection{Modifications to Einstein's equation}

The modifications to the Einstein equation can now be found by substituting equations \eqref{eqn:varcG2} and \eqref{varPan2} into equation \eqref{varRiegert} to evaluate
the contribution to the right-hand side of equation \eqref{MEFT}. 
The first term in equation \eqref{varRiegert}, gives the following modification 
at point $x$ (with free indices $\al,\beta$):
\begin{align}
-\frac23& \int d^4 y \sqrt{-g(x)} \mathcal{G}(y)  \Big[2 \left( g_{\al \b} \Box - \na_{\al} \na_{\b} \right)\Box  +3 \big( 4 R_{\mu \al\b \nu} + 4 R_{\mu \nu} g_{\al\b} - 8 R_{\mu \al} g_{\nu \b}   \notag\\[2mm]
  & \; + \frac{14}{3} R_{\al \b} g_{\mu \nu} - 2 R g_{\al \b} g_{\mu \nu}+ 2 R g_{\al \mu} g_{\b \nu} \big) \na^{\mu} \na^{\nu} +  \Big( \na^{\mu} R g_{\al \b} - 2 \de^{\mu}_{\al} \na_{\b} R \Big) \na_{\mu} 
  \Big] \GPan (x,y), \label{1stvarcR}
\end{align}
while the second variation in equation \eqref{varRiegert}, gives
\begin{align}
 -& \int d^4 y \int d^4 z \sqrt{-g(x)} \mathcal{G}(y) \mathcal{G}(z)   \Big[\frac23 \na_{\mu}\na_{\al} \na_{\b}\GPan(x,y) \na^{\mu}\GPan(x,z)   \notag\\[3mm] 
& \hspace{25mm}- 2\na_{(\al} \Box\GPan(x,y) \na_{\b)}\GPan(x,z) + \frac13 g_{\al \b} \na_{\mu} \Box \GPan(x,y) \na^{\mu}\GPan(x,z)
 \notag\\[3mm]
& \hspace{25mm}+  2 \na_{\al} \na_{\b} \GPan (x,y)\Box \GPan(x,z) - \frac12 g_{\al \b} \Box \GPan(x,y)\Box \GPan(x,z) 
    \notag \\[3mm]
 &\hspace{25mm} - \frac43 \na^{}_{\al}\na^{\mu} \GPan(x,y) \na_{\b} \na_{\mu}\GPan(x,z)  + \frac13 g_{\al \b} \na^{\mu} \na^{\nu} \GPan(x,y) \na_{\mu} \na_{\nu}\GPan(x,z)  \notag \\[3mm]
 & \hspace{25mm}- 4 R_{\mu (\al} \na_{\b)} \GPan(x,y) \na^{\mu} \GPan(x,z) - \frac43 R_{\mu \al \nu\b }\na^{\mu}  \GPan(x,y)  \na^{\nu} \GPan(x,z) \notag \\[3mm]
 &\hspace{25mm} + \frac43 g_{\al \b} R_{\mu \nu} \na^{\mu} \GPan(x,y) \na^{\nu} \GPan(x,z) + \frac23 R \na_{\al} \GPan(x,y) \na_{\b} \GPan(x,z) \notag \\[3mm]
&\hspace{25mm}+ \frac23  R_{\al \b} \na^{\mu} \GPan(x,y) \na_{\mu} \GPan(x,z) - \frac13 g_{\al \b} R \na^{\mu} \GPan(x,y) \na_{\mu} \GPan(x,z) \Big]. \label{2ndvarcR}
\end{align}
In the two equations above, the differential operators are all with respect to the coordinate $x$.  
Since the Riegert action is covariant, the modifications to the Einstein equation must be divergenceless. Using the Schouten identity
\begin{equation}
 4 R_{\mu \nu \al \b } R^{\tau \b \mu \nu} \na_{\tau} = E_4 \na_{\al} - 4 R R_{\al \b} \na^{\b} + 8 R_{\al \b \mu\nu} R^{\b \nu} \na^{\mu} + 8 R_{\al \b} R^{\b \mu} \na_{\mu},
\end{equation}
it can be shown that the divergence of the first contribution, \eqref{1stvarcR}, is 
\begin{equation}
 - \int d^4 y \,  \mathcal{G}(x) \mathcal{G}(y) \na_{\al} \GPan (x,y).
\end{equation}
Moreover, the divergence of the second contribution, \eqref{2ndvarcR}, is 
\begin{equation}
 \frac12 \int d^4 y \int d^4 z \, \mathcal{G}(y) \mathcal{G}(z) \left( \na_{\al} \GPan(x,z) \Pan (x) 
 \GPan (x,y)  + \na_{\al} \GPan (x,y) \Pan(x) \GPan (x,z) \right),
\end{equation}
which by definition of the Green function, \eqref{Pan1}, is equal and opposite to the divergence of the first contribution. Therefore, the modification of the Einstein equation produced by the Riegert action is indeed divergenceless. This reassures us of the consistency of our equations of motion.

The expressions \eqref{1stvarcR} and \eqref{2ndvarcR} appear on the right-hand side 
of the Einstein equation, \eqref{MEFT}. For a spatially flat universe, the FRW metric takes the conformally flat form \eqref{ds} where $\eta$ is the conformal time, and where for 
concreteness we specialise to the physically relevant case of the radiation era (\ref{aeta}). 
The FRW metric is clearly supplemented by inhomogeneities and anisotropies but 
we choose to work in a simple setting where we can study precisely how the new 
corrections to the Einstein equation affect the background (taking into account 
inhomogeneities would anyhow  not alter our general conclusions).
We will not try to solve the highly complicated system of integrodifferential equations 
to determine what the new FRW-like solution is -- if there is any solution -- 
but we will simply ask how large the backreaction of the new terms on the 
background geometry is. Moreover, we look only at the $\eta \eta$ components of the 
Einstein equation -- the other components receive contributions of the same order.

Using the equations in appendix \ref{app:conf}, the contribution to the equation of motion at point $x= (\eta, \bx)$ coming from the variation of Green function is 
\begin{align} \label{vargreeneom}
   \frac{(24)^2}{3} &\int \frac{ d^4 y}{\eta^4_y} \int \frac{d^4 z}{\eta^4_z} \Big[ 2 \na_{\eta}\na_{\eta}\na_{\eta} \GPan(x,y) \na_{\eta} \GPan(x,z) - 2\na_{i}\na_{\eta}\na_{\eta} \GPan(x,y) \na_{i} \GPan(x,z) \notag \\[3pt]
  & + 5 \na_{\eta}\Box \GPan(x,y) \na_{\eta} \GPan(x,z) + \na_{i}\Box \GPan(x,y) \na_{i} \GPan(x,z) \notag  - \frac32 \Box \GPan(x,y) \Box \GPan(x,z) \\[7pt]
  &-6 \na_{\eta}\na_{\eta} \GPan(x,y) \Box \GPan(x,z)- 3 \na_{\eta}\na_{\eta} \GPan(x,y) \na_{\eta}\na_{\eta} \GPan(x,z) \notag \\[7pt] 
  & +2 \na_{i}\na_{\eta} \GPan(x,y) \na_{i}\na_{\eta} \GPan(x,z) + \na_{i}\na_{j} \GPan(x,y) \na_{i}\na_{j} \GPan(x,z)  \notag \\[7pt]
  &- \frac{18}{\eta^2} \na_{\eta} \GPan(x,y) \na_{\eta} \GPan(x,z) + \frac{2}{\eta^2} \na_{i} \GPan(x,y) \na_{i} \GPan(x,z) \Big],
\end{align}
where the derivative operators are with respect to the coordinate $x.$
The factors of $1/\eta^4_y$ and $1/\eta^4_z$ come from $\mathcal{G}(y)$ and $\mathcal{G}(z)$, evaluated in the radiation era. It is the integration of these factors to the beginning of the radiation era that produces the large effects.
Furthermore the contribution from the variation of $\mathcal{G}(y)$ and $\mathcal{G}(z)$ is
\begin{align}
- 48 &\int d^4 y \frac{1}{\eta^4_y}  \left[ \frac23 \left(\Box + \na_{\eta}\na_{\eta} \right) \Box \GPan(x,y) - \frac{14}{\eta^2} \Box \GPan(x,y) - \frac{12}{\eta^2} \na_{\eta}\na_{\eta}  \GPan(x,y)\right].
\end{align}

The integrals can now be evaluated. For example,
\begin{align}
 \int d^4 y \frac{1}{\eta^4_y} \na_{\eta}  \GPan(x,y)&= \int_{L}^{\etarad} d \eta_y \cdot 4 \pi 
 \int dy y^2 \frac{1}{\eta^4_y} \cdot \frac{1}{8 \pi}\delta(\eta- \eta_y - |\by|) \notag \\ 
 %&= \frac{1}{2} \int_{L}^{\etarad} d \eta_y  \frac{(\eta - eta_y)}{\eta_y^4} \\
 &= \frac{\etarad^2}{6 L^3} - \frac{\etarad}{2 L^2}+ \frac{1}{2 L} - \frac{1}{6 \etarad}. \label{edG}
\end{align}
It is clear from the above integration that the $1/L^3$ effect is coming from integrating the 
factor $1/\eta^4_y$ back to $L$. Evaluating the other integrals, we have the following $1/L$ 
dependence (always integrating from $L$ to $\etarad$)
\begin{gather}
  \int d^4 y \frac{1}{\eta^4_y} \Box  \GPan(x,y) = - \frac{2 \etarad}{3 L^3} + \frac{3}{2 L^2}
  - \frac{1}{\etarad L}, \notag\\[5pt]
\int d^4 y \frac{1}{\eta^4_y}  \na_{\eta}\na_{\eta}  \GPan(x,y) = 
 \frac{\etarad}{6  L^3} - \frac{1}{2 \etarad L}, \notag \\[5pt]
\int d^4 y \frac{1}{\eta^4_y}  \na_{i}\na_{j}  \GPan(x,y) = \left( - \frac{\etarad}{6 L^3} + \frac{1}{2 L^2} - \frac{1}{2 \etarad L} \right)\de_{ij}, \notag \\[5pt]
     \int d^4 y \frac{1}{\eta^4_y}  \na_{\eta}\Box  \GPan(x,y) = \frac{2}{3 L^3} - \frac{3}{\etarad L^2}+ \frac{3}{\etarad^2 L} ,   \notag\\[5pt]
 \int d^4 y \frac{1}{\eta^4_y}  \na_{\eta}\na_{\eta}\na_{\eta}   \GPan(x,y) = - \frac{1}{6 L^3} + \frac{3}{2 \etarad^2 L}, \;    \notag \\[5pt]
 \int d^4 y \frac{1}{\eta^4_y} \Box^2  \GPan(x,y) =
  - \frac{3}{\etarad^2 L^2} + \frac{6}{\etarad^3 L}, \notag \\[5pt]
   \int d^4 y \frac{1}{\eta^4_y}  \na_{\eta}\na_{\eta}  \Box  \GPan(x,y) = - \frac{2}{\etarad L^3} + \frac{12}{\etarad^2 L^2} - \frac{15}{\etarad^3 L}.
\label{leading}
\end{gather}
It is clear that the leading contribution comes from the variation of the Green's function, expression \eqref{vargreeneom}. Substituting the above values for the integrals into expression \eqref{vargreeneom} we find
\begin{equation}                                                                                                                                                                                                          
  - \frac{24}{L^4}.                                                                                                                                                                                             \end{equation}
Quite remarkably, all other dependence on $1/L$ cancels (this feature may be related to
the fact that in lowest non-trivial order, one of the $\Box^{-1}$ in $\GPan$ cancels \cite{DS}).
In particular, the leading and subleading terms cancel ($1/L^6$ and $1/L^5$). However,  from equation (\ref{L}) we see that we can, nevertheless, take $n_*$ rather large and still get a sizable effect.

Equation \eqref{edG} illustrates that the $1/L^3$ effect comes from the $1/\eta_y^4$ in the integral which is then integrated back to the beginning of the radiation era, $L$. This factor comes from 
the $\mathcal{G} (y)$ in the integral, more precisely the Riemann-squared term (see appendix \ref{app:conf}) and the convolution integral
\begin{equation} \label{leadconv}
\GPan \star  R_{\mu \nu \rho \sigma} R^{\mu \nu \rho \sigma} \, .
\end{equation}
(because $C_{\mu\nu\rho\si}=0$ we have $R_{\mu\nu} R^{\mu\nu} = - \frac12
R_{\mu\nu\rho\si}R^{\mu\nu\rho\si}$).
It is also clear from equations \eqref{leading} that the number of derivatives acting on the above convolution, generically, does not change the leading order behaviour -- the exception is 
the $\Box^2$ derivative or, by spherical symmetry, single spatial derivatives.
Note also that in ref.~\cite{Mott1} it was precisely the other term $\propto \Box R$ in $\mathcal{G}$
which played the dominant role, but is negligible for our analysis.

The leading behaviour of the convolution \eqref{leadconv} is also independent of whether 
the Green's function of the Paneitz operator or the second-order conformally invariant 
operator is used, as in ref.~\cite{MN}, see also following section. We, therefore, find that 
the effect observed in ref.~\cite{MN} appears to be rather generic and will be present 
whenever the Einstein equations are modified by terms that include convolutions 
of the form \eqref{leadconv}, where $\GPan$ can also be replaced by the Green's 
function $G$, \eqref{G2}, defined by  equation \eqref{Green1}.

\section{Local Action with Dilaton} \label{sec:local}

 When conformal symmetry is spontaneously broken, there is a Goldstone boson, 
 the dilaton $\tau$, which, under Weyl transformations (see appendix \ref{app:weyl}), transforms as
 \begin{equation}
  \tau \rightarrow \tau + \sigma.
 \end{equation}
The dilaton can be used to write a {\em local} action which gives the conformal anomaly upon 
variation. In particular
it was shown in ref.~\cite{ST} that this action agrees with the local form of the Riegert action 
only up to cubic order, but not beyond. As in the case of unbroken conformal symmetry, the action is not fixed uniquely 
by the requirement that it reproduces the anomaly. The correct action is constrained 
by Ward identities, which are satisfied because the symmetry is only broken spontaneously, and analyticity properties, namely the existence of poles that correspond to the dilaton in tree diagrams~\cite{ST}. The anomalous action in the unbroken phase will have a different analytic structure. We will now show that upon elimination of the dilaton by its equation of motion there will be enhancement effects similar to the ones 
exhibited in the foregoing section.

In the spontaneously broken phase, the anomalous local action is~\cite{ST}
\begin{align} \label{ST}
W = & - a \int d^4 x \sqrt{-g} \left[ \frac1f \tau E_4 + \frac{2}{f^2} G^{\mu\nu}
\partial_{\mu} \tau \partial_{\nu} \tau + \frac{4}{f^3} \partial^{\mu} \tau \partial_{\mu} \tau \Box \tau - \frac{2}{f^4} (\partial^{\mu} \tau \partial_{\mu} \tau)^2 \right] \notag \\[3pt]
 & - c \int d^4 x \sqrt{-g} \tau C^{\mu \nu \rho \sigma} C_{\mu \nu \rho \sigma},
\end{align}
where $G^{\mu\nu} \equiv  R^{\mu \nu} - \frac12 R g^{\mu \nu} $ is the Einstein
tensor, and $f$ sets the scale of conformal symmetry breaking, with $\langle\vp\rangle = f$.
The Weyl transformation property of the dilaton is such that the $\tau$ variation of the first terms in first and second line gives the anomaly, while the rest of the terms on the first line correct for the 
variation of the Euler term under Weyl transformations. These terms are not required for 
the Weyl-squared density as it is Weyl invariant. In ref.~\cite{ST}, it was shown that this 
action has the correct properties expected of the anomalous action when conformal 
symmetry is spontaneously broken.

The action \eqref{ST} is evidently local, hence one might expect that the effects, derived in section \ref{sec:riegert}, that are a consequence of the non-local Riegert action will not persist. However, the Riegert action can itself be written as a local action by introducing an auxiliary scalar field, 
the `conformalon' (that action is different from (\ref{ST})). Therefore, the fact that the 
action is local does not necessarily mean that the Einstein equation will not receive large 
corrections. In particular, the `conformalon' can encode the effects of the non-local terms. 
In this section, we analyse the behaviour of the dilaton $\tau$ in action~(\ref{ST}) and show that it encodes effects similar to the ones seen in the non-local Riegert action.

Following ref.~\cite{ST}, we supplement the anomalous action \eqref{ST} by a quadratic action for the dilaton, 
\begin{equation}
 S_0 = - \int d^4 x \sqrt{-g} \left( \partial^{\mu} \varphi \partial_{\mu} \varphi + \frac16 R (\varphi - f)^2 \right),
\end{equation}
where the fields $\vp$ and $\tau$ are related by
\begin{equation}
 \varphi = f\big(1 - e^{-\tau/f} \big) \; \Rightarrow \;\; \quad \vp - f \rightarrow e^{- \sigma/f} (\varphi -f).
\end{equation}
Since $ (\varphi -f)$ is a scalar of conformal weight $-1$, this is simply the action for a 
conformally coupled scalar, that is, the action $S_0$ is Weyl invariant. 

The dilaton can be integrated out perturbatively to give an infinite series for the 
anomalous effective action~\cite{ST}. Let
\begin{equation}
 S= S_0 + \epsilon W, \qquad \varphi= \varphi_0 + \sum_{n\geq 1} \epsilon^n \varphi_{n},
\end{equation}
where 
\begin{equation}\label{vp0}
\varphi_0(x) = \frac{f}{6} \int d^4 y \sqrt{-g} \, G(x,y) R(y) 
\equiv  \frac{f}{6} (G \star R)(x)
\end{equation}
is found by extremising the $S_0$ action, and $G(x,y)$ is the Green's function 
defined in (\ref{Green1}) for the conformal d'Alembertian; the
symbol $\star$ is shorthand for the convolution integral. We have also introduced 
$\epsilon$ as a bookkeeping parameter for the perturbative expansion--- this parameter 
is not necessarily small, and is set to unity after the calculation. 

Integrating out the dilaton, the action is given by 
\begin{equation}
 S= \sum_{n \geq 0} \epsilon^n S^{(n)},
\end{equation}
where
\begin{align}
 S^{(0)} &= S_0 \big|_{\varphi_0}, \qquad
 S^{(1)} = W \big|_{\varphi_0}, \\[2mm]
 S^{(2)} &= -  \frac12\left( \frac{1}{\sqrt{-g} } \frac{\de W}{\de \varphi} \star G \star \frac{1}{\sqrt{-g} } \frac{\de W}{\de \varphi} \right)\bigg|_{\varphi_0}, \\[2mm]
 S^{(3)} &= -  \frac12 \frac{1}{\sqrt{-g} } \frac{\de^2 W}{\de \varphi^2} \star\left( G \star  \frac{1}{\sqrt{-g} }  \frac{\de W}{\de \varphi} \right)^2\bigg|_{\varphi_0}, 
\end{align}
and so on. The solution for the dilaton can be expanded as
\begin{equation}\label{vp}
 \varphi= \varphi_0 - \epsilon \,  G \star \frac{1}{\sqrt{-g} } \frac{\de W}{\de \varphi} \bigg|_{\varphi_0} - \epsilon^2 G  \star \left(  \frac{1}{\sqrt{-g} } \frac{\de^2 W}{\de \varphi^2} G \star \frac{1}{\sqrt{-g} } \frac{\de W}{\de \varphi} \right) \bigg|_{\varphi_0} + \dots 
\end{equation}
To estimate the deviation of $\vp$ from the free field solution $\vp_0$ in \eqref{vp0}, we observe that
\beq
\frac{\de W}{\de \vp} = \frac1{1 - \vp/f} \frac{\de W}{\de \tau}
\eeq
and
\beq
\frac{\de^2 W}{\de \vp^2} = \frac1{(1 - \vp/f)^2} \left( \frac1{f} \frac{\de W}{\de \tau}
    \,+\,  \frac{\de^2 W}{\de \tau^2} \right),
\eeq
and so on. Because $\tau\sim 0$ corresponds to $\vp\sim 0$, we can ignore the
prefactor in this approximation; furthermore
\beq
\frac1{\sqrt{-g}} \frac{\de W}{\de\tau} = -\frac{a}{f} \, \GB \,+\, \cO(\tau) \;,\quad
\frac1{\sqrt{-g}} \frac{\de^2 W}{\de\tau^2} = \cO(\tau)
\eeq
Returning to the expansion (\ref{vp}) we see that for consistency all higher order terms
would have to remain small in comparison with the lowest order solution (\ref{vp0}). 
While this is certainly true for the weak field expansion, we
will now show that this need not be the case in a general cosmological setting.

To determine the induced corrections, we first observe that the modifications to Einstein's
equations are simply given by the dilaton energy momentum tensor
\bea\label{Tphi}
T_{\mu\nu}^{(\vp)} &=& - \na_\mu \vp \na_\nu \vp + \frac12 g_{\mu\nu} (\pa \vp)^2 
+ \frac1{12} g_{\mu\nu} R (\vp-f)^2 - \frac16 R_{\mu\nu} (\vp-f)^2 \nn\\[2mm]
&& - \frac16 g_{\mu\nu} \Box\big((\vp-f)^2\big) + \frac16 \na_\mu\na_\nu \big((\vp-f)^2\big)
\,+\, \epsilon \frac1{\sqrt{-g}} \frac{\de W}{\de g^{\mu\nu}}
\eea
which is indeed traceless by the lowest order equations of motion if one ignores the last 
term on the right-hand side. To analyze 
the effect of the non-localities on any given solution of the classical Einstein equations, we 
simply substitute the solution for the dilaton field in this given background. For consistency 
and in accordance with our assumptions the dilatonic correction then should remain very 
small (in which case a linearized treatment of the full set of equations is justified). We will 
now show that in general this is not the case.

As for the Riegert action, we consider a cosmological solution of the classical 
Einstein equations with conformally flat metric (\ref{ds}), corresponding to the radiation 
era ending at $\eta\!=\!\etarad$ and starting at $\eta = L$, with the linear dependence 
described by (\ref{aeta}). Then $R=0$, so with (\ref{vp0}), $\vp_0$ receives contributions 
only from primordial perturbations, which can be crudely estimated as 
\beq
\vp_0 \sim 10^{-5} f  \quad \Rightarrow \quad   \pa_\eta \vp_0 \sim 10^{-5} \frac{f}{\etarad}
\eeq
The corresponding contribution to the energy momentum tensor at zeroth order 
is thus very small, of order $10^{-10} f^2 \etarad^{-2}$, and by itself would not
lead to any observable effect. However, the $\cO(\epsilon)$ term 
in (\ref{vp}) receives a dramatically bigger contribution since  in the radiation era 
$\GB \propto \etarad^4/\eta^{-8}$. Thus,
\bea
\vp_1(\eta) &\propto& \frac{\etarad^4}{f}
\int\!\rd^3 y \! \int^{\eta}_{L}\!\!\rd\eta'\, \frac{a(\eta')^4}{4\pi |\bx - {\bf y}|}
\frac{\de(\eta-\eta'-|\bx - {\bf y}|)}{a(\eta)a(\eta')}\frac1{(\eta')^8} \nn\\[2mm]
&=& \frac{\etarad^2}{f \eta}  \int_{|\by| < \eta - L} d^3y  \, 
\frac1{4\pi |\by|}\cdot \frac1{(\eta - |\by|)^5} \nn\\[2mm]
&=& \frac{\etarad^2}{f \eta} \left[ - \frac13 \frac1{L^3} \,+\,  \frac14 \frac{\eta}{L^4} 
      \, + \, \frac1{12} \frac1{\eta^3} \right]
\eea
where in the second line we have shifted the integration variable $\bf$ so as to
make explicit the independence of $\vp_1$ on the spatial coordinates (this is only
due to our simplifying choice of a homogeneous and isotropic background; if one 
takes into account spatial inhomogeneities would not alter our main conclusions).
Differentiating and setting $\eta=\etarad$  we see that
\beq
\pa_\eta \vp_1 = \frac13 \frac1{f L^3} + \cO\left( \frac1{f\etarad^3} \right)
\eeq
thus recovering the result from the previous section, except that $\etarad$
is replaced by the inverse dilaton coupling $f^{-1}$ (which, incidentally, shows that
this analysis becomes invalid in the limit $f\rightarrow 0$ when 
conformal symmetry is restored and the action (\ref{ST}) is no longer
the correct one). It is thus clear that there is again a substantial enhancement even if 
$L$ is much larger than the Planck time $\ePL$, as in the foregoing section.
It is now straightforward to determine the correction to the Einstein equations,
for which we need only the first derivative $\pa_\eta \vp_1$, by substituting 
the above result into (\ref{Tphi}). Inspection of the higher order terms in (\ref{vp})
shows that these will produce similarly large contributions. In conclusion,
the series expansion (\ref{vp}) will diverge due to large contributions from
the non-localities.

\vspace{1cm}

\section{Conclusions}

In this paper we have assembled evidence that the inclusion of an anomalous action in the effective field theory can lead to significant effects arising from the non-local nature of the action. In the local versions, the effect is encoded in the behaviour of the conformalon or dilaton field. This effect is fairly generic and arises from an integration over the convolution of the Green function of some conformally invariant differential operator and the Riemann tensor over a range where the curvature can be large, but still remaining outside the Planck regime. 

We exhibit this feature for the non-local Riegert action, which is the analogue of the Polyakov action in four dimensions. Even though the action is known not to possess the correct analyticity structure, it is given in closed form and hence we can explicitly demonstrate this effect. The correct action in the regime of unbroken conformal symmetry is not known. However, given that conformal symmetry is badly broken in nature, the anomalous action for spontaneously broken conformal symmetry, which is known in the literature, is of greater interest. We study the aforementioned action and show that the effects found in the case of the Riegert action reappear in the expansion for the dilaton, invalidating the expansion. It is possible (though not likely) that this expansion can be resummed to remove this effect and we have not excluded this possibility in the paper.

\vspace{3mm}

\noindent{\bf {Acknowledgments:}}  K.A.M. would like to thank the AEI for hospitality and 
support during this work; he was supported by the Polish NCN grant 
DEC-2013/11/B/ST2/04046. H.N. would like to thank J.~Erdmenger, Y.~Gusev, H.~Osborn and  
A. Schwimmer for discussions and correspondence.

 \appendix
 
 \section{Variations} \label{app:variations}

In this appendix, we collect the variation of some tensors with respect to the metric. These variations are used to derive the corrections to the Einstein equation from the effective action.

The variation of the Christoffel connection and the curvature tensor is
\bea
\de \Ga^\rho_{\mu\nu} &=&    \frac12 g^{\rho\la} \big( \nabla_\mu \de g_{\la\nu}
    + \nabla_\nu \de g_{\la\mu} - \nabla_\la \de g_{\mu\nu} \big),               \nn\\[2mm]
\de R^\rho{}_{\si\mu\nu} &=& 
\frac12 g^{\rho\la} \big[ \nabla_\mu , \nabla_\nu \big] \de g_{\la\si} +\, g^{\rho\la} \big( \nabla_\mu \nabla_{[\si} \de g_{\la]\nu} \,-\,
            \nabla_\nu \nabla_{[\si} \de g_{\la]\mu} \big),
\eea
which immediately gives
\bea
%\de R_{\mu\nu\rho\si}&=&\na_\mu\na_{[\si}\de g_{\rho ]\nu}-\na_\nu\na_{[\si}\de g_{\rho ]\mu}\\
%&&+\frac12 R^{\al}{}_{\nu\rho\si}\de g_{\al\mu}-\frac12 R^{\al}{}_{\mu\rho\si}\de g_{\al\nu}\nn\\
\de R_{\mu\nu}&=&-\frac12 \Box\de g_{\mu\nu}-\frac12 \na_\mu\na_{\nu}(g^{\al\beta}\de g_{\al\beta})+\frac12\na^\la \na_\mu \de g_{\nu\la}+\frac12\na^\la \na_\nu \de g_{\mu\la}\nn\\
\de R &=&-R^{\al\beta}\de g_{\al\beta}-\Box(g^{\al\beta}\de g_{\al\beta})+\na^\mu\na^{\nu}\de g_{\mu\nu}\nn
\eea

From above, we can write
\begin{align}
\frac{1}{\sqrt{-g}}\de \big(\sqrt{-g}R_{\mu\nu\rho\si}R^{\mu\nu\rho\si}\big) &= 
\left[4 R^{\mu\al\beta\si}\na_\mu\na_\si  -2(R^{\al\nu\rho\si}R^{\beta}{}_{\nu\rho\si}-\frac14 R_{\mu\nu\rho\si}R^{\mu\nu\rho\si}g^{\al\beta})\right]\de g_{\al\beta}\nn\\[2mm]
\frac{1}{\sqrt{-g}}\de \big(\sqrt{-g}R_{\mu\nu}R^{\mu\nu}\big) &= \left[2 R^{\mu\beta}\na^\al \na_\mu - g^{\al\beta} R^{\mu\nu}\na_\nu\na_\mu-R^{\al\beta}\Box \right. \notag\\
& \hspace{35mm} \left. -2(R^{\mu\al}R^{\nu\beta}-\frac14R^{\mu\nu}R_{\mu\nu}g^{\al\beta})\right]\de g_{\al\beta}\nn\\[3mm]
\frac{1}{\sqrt{-g}}\de\big(\sqrt{-g}R^2\big) &= \left[2R(\na^\al\na^\beta- g^{\al\beta}\Box)
-2R(R^{\al\beta}-\frac14 R g^{\al\beta})\right]\de g_{\al\beta}
\end{align}
It is easy to see that the only combination that does not depend on the trace part is $\sqrt{-g}C^2=\sqrt{-g}(R_{\mu\nu\rho\si}R^{\mu\nu\rho\si}-2R_{\mu\nu}R^{\mu\nu}+\frac13 R^2)$.

The variation of the box operator, when acting on a scalar quantity, is
\begin{equation}
 \frac{1}{\sqrt{-g}} \de ( \sqrt{-g} \Box) = \na_{\alpha} \left[ \left( \de g^{\alpha \beta} - \frac12 g^{\alpha \beta} g_{\gamma \de} \de g^{\gamma \de} \right) \partial_{\beta} \right].
\end{equation}
Therefore,
\begin{equation}
 \frac{1}{\sqrt{-g}} \de ( \sqrt{-g} \Box R ) = \Box \left[ \left( R_{\al \b} + g_{\al \b} \Box - \na_{\al} \na_{\b} \right) \de g^{\al \b} \right] + \na_{\alpha} \left[ \left( \de g^{\alpha \beta} - \frac12 g^{\alpha \beta} g_{\gamma \de} \de g^{\gamma \de} \right) \partial_{\beta} R \right]
\end{equation}

\section{Weyl transformations} \label{app:weyl}

We collect a list of the transformation of some tensors under Weyl transformation 
\begin{equation}
 g_{\mu \nu} \longrightarrow \Omega^2 \, g_{\mu \nu} = e^{2 \sigma } \, g_{\mu \nu} .
\end{equation}
The curvature tensor,
\begin{equation}
 R_{\mu \nu \rho \sigma} = C_{\mu \nu \rho \sigma} + 2 g_{\mu[\rho} \, P_{\sigma]\nu} - 2 g_{\nu[\rho} \, P_{\sigma]\mu},
\end{equation}
its contractions and the Schouten tensor,
\begin{equation}
 P_{\mu \nu} = \frac{1}{2} \left( R_{\mu \nu} - \frac{1}{6} \,g_{\mu \nu}\, R \right),
\end{equation}
transform as follows:
\begin{align}
 R^{\mu}{}_{\nu \rho \sigma} \longrightarrow & \, R^{\mu}{}_{\nu \rho \sigma} - 2 \,\delta^{\mu}_{[\rho} \, \nabla^{}_{\sigma]} \nabla_{\nu} \sigma + 2 \,g^{\mu \al}\, g_{\nu[\rho}\, \nabla_{\sigma]} \nabla_{\al} \sigma + \delta^{\mu}_{[\rho} \, \partial^{}_{\sigma]} \sigma \, \partial_{\nu} \sigma \notag\\[3pt]
 &  - 2 \, g^{\mu \al} \,g_{\nu[\rho}\, \partial_{\sigma]} \sigma \,  \partial_{\al} \sigma - 2 \, \delta^{\mu}_{[\rho}\, g^{}_{\sigma] \nu}\, g^{\al \b}\, \partial_{\al} \sigma \, \partial_{\b} \sigma.\\[3pt]
 R_{\mu \nu} \longrightarrow & \,R_{\mu \nu} - 2\, \nabla_{\mu} \nabla_{\nu} \sigma + 2\, \partial_{\mu} \sigma \, \partial_{\nu} \sigma - g_{\mu \nu} \, \Box \sigma - 2 \, g_{\mu \nu}\, g^{\rho \sigma}\, \partial_{\rho} \sigma \, \partial_{\sigma} \sigma, \\[3pt]
 R \longrightarrow & \, \Omega^{-2} \left( R - 6 \, \Box \sigma  - 6 \, g^{\mu \nu}\, \partial_{\mu} \sigma \, \partial_{\nu} \sigma \right), \\[3pt]
 P_{\mu \nu} \longrightarrow & \,P_{\mu \nu} - \, \nabla_{\mu} \nabla_{\nu} \sigma + \, \partial_{\mu} \sigma \, \partial_{\nu} \sigma  - \frac{1}{2} \, g_{\mu \nu}\, g^{\rho \sigma}\, \partial_{\rho} \sigma \, \partial_{\sigma} \sigma.
  \end{align}
The covariant derivative also transforms under a Weyl transformation. In particular, the Christoffel symbol transforms as 
\begin{equation}
 \Gamma^{\rho}_{\mu \nu} \longrightarrow \Gamma^{\rho}_{\mu \nu} + 2 \, \delta^{\rho}_{(\mu} \, \partial^{}_{\nu)} \sigma - g^{\rho \sigma} \, g_{\mu \nu} \, \partial_{\sigma} \sigma.
\end{equation}
For completeness, the spin connection transforms as
\begin{equation} \label{spincontrans}
 \omega_{\mu}{}^{\al \b} \longrightarrow  \omega_{\mu}{}^{a b} + 2 e_{\mu}^{[a} e_{\nu}^{b]} g^{\nu \rho} \partial_{\rho} \sigma.
\end{equation}

\section{Weyl invariant actions for spins $s=0,\frac12, 1$}

Given the transformation property of the quadratic operator introduced in equation \eqref{Green1},
\begin{equation}
 \sqrt{-g} \left( - \Box + \frac16 R \right) \longrightarrow \Omega^2 \sqrt{-g} \left( - \Box + \frac16 R \right) - \Omega^2 g^{\mu \nu} \left(\partial_{\mu} \sigma \partial_{\nu}\si
  + \partial_{\mu} \sigma \partial_{\nu}  \right) - \Omega^2 \Box\si
\end{equation}
the operator is Weyl covariant if it acts on a scalar $\phi$ of conformal weight $-1$,
\begin{equation}
  \phi \longrightarrow \Omega^{-1} \phi.
\end{equation}
Furthermore, it is then clear that
\begin{equation}
  \sqrt{-g} \phi \left( - \Box + \frac16 R \right) \phi 
\end{equation}
is Weyl invariant.

For a spinor $\psi$ of conformal weight $-\frac32$, 
\begin{equation}
  \psi \longrightarrow \Omega^{-\frac32} \psi,
\end{equation}
the Dirac action 
\begin{equation}
  \overline{\psi} \gamma^{\mu} \nabla_{\mu} \psi   \equiv
  \overline{\psi} \gamma^{\mu}  \left( \partial_\mu + \frac14 \omega_{\mu\,ab} \ga^{ab}\right)\psi,
\end{equation}
 is Weyl-invariant without any modification. This can be seen using the the transformation of the spin connection, \eqref{spincontrans}, and noting that 
 \begin{equation}
  \gamma_{\mu} \gamma^{\mu \nu} = 3 \gamma^{\nu}.
 \end{equation}
The invariance of the Yang-Mills action is anyhow clear because of the invariance
of the factor $\sqrt{-g} g^{\mu\rho} g^{\nu\si}$ multiplying 
${\rm Tr}( F_{\mu\nu} F_{\rho\si})$ under Weyl transformations.

\section{Quantities in conformally flat spacetime} \label{app:conf}

In conformally flat spacetime (where $C_{\mu\nu\rho\si} = 0$) the metric can be chosen to be
\begin{equation}
 g_{\mu \nu} dx^\mu dx^\nu = a(\eta)^2 ( -d\eta^2 + d\bx^2)
\end{equation}
With respect to this metric,
\begin{gather}
 \Gamma^{\eta}_{\eta \eta} = \frac{a'}{a}, \qquad \Gamma^{\eta}_{ij} = \frac{a'}{a} \delta_{ij}, \qquad \Gamma^{i}_{\eta j} = \frac{a'}{a} \delta^{i}_{j}, \\[3mm]
 R^{\eta}{}_{i \eta j} = \left( \frac{a'}{a} \right)' \delta_{ij}, \qquad R^{i}{}_{\eta j \eta} =- \left( \frac{a'}{a} \right)' \delta^{i}_{j}, \qquad R^{i}{}_{j k l} = \left(\frac{a'}{a} \right)^2 \left( \delta^{i}_{k} \delta_{jl} - \de^{i}_{l} \de_{jk} \right),\\[3mm]
 R_{\eta \eta} = -3 \left( \frac{a'}{a} \right)', \qquad R_{ij} = \left( \frac{a''}{a} + \left( \frac{a'}{a} \right)^2 \right) \delta_{ij}, \qquad R = 6\,  \frac{a''}{a^3},
\end{gather}
where the prime denotes differentiation with respect to $\eta$.


\begin{thebibliography}{99}
\bibitem{MN} K.A.~Meissner and H.~Nicolai, {\tt arXiv:1607.07312 [gr-qc]}
\bibitem{Riegert} R.J.~Riegert, Phys. Lett. {\bf B134} (1984) 56
\bibitem{ST} A.~Schwimmer and S.~Theisen, Nucl. Phys. {\bf B847} (2011) 590
\bibitem{KS} Z. Komargodski and A.~Schwimmer, JHEP {\bf 1112} (2011) 099
\bibitem{Deser2} S.~Deser, M.J.~Duff and C.~Isham, Nucl. Phys. {\bf B111}(1976) 45.
\bibitem{Duff1} M.J.~Duff, Nucl. Phys. {\bf B125}(1977) 334
\bibitem{CD} S.M. Christensen and M.J. Duff, Phys. Lett. {\bf 76B} (1978) 571.
\bibitem{FT} E. Fradkin and A. Tseytlin,  Nucl.Phys. {\bf B203}(1982) 157
\bibitem{DS}S.~Deser and A.~Schwimmer, Phys. Lett. {\bf B309} (1993) 279
\bibitem{EO} J.~Erdmenger and H.~Osborn, Nucl. Phys. {\bf B483}(1997) 431
\bibitem{Deser1} S.~Deser,  Helv.Phys.Acta {\bf 69} (1996) 570;
       Phys. Lett. {\bf B 479} (2000) 315       
\bibitem{Duffrev} M.~J.~Duff,  %``Twenty years of the Weyl anomaly,''
  Class.\ Quant.\ Grav.\  {\bf 11}, 1387 (1994)
  %doi:10.1088/0264-9381/11/6/004
  [hep-th/9308075].
\bibitem{BGVZ} A.O.~Barvinsky, Y.V. Gusev, G.A.~Vilkovisky and V.V.~Zhytnikov,
        Nucl. Phys {\bf B439} (1995) 561
\bibitem{Barvinsky:1995it} 
  A.~O.~Barvinsky, A.~G.~Mirzabekian and V.~V.~Zhytnikov,
  %``Conformal decomposition of the effective action and covariant curvature expansion,''
  gr-qc/9510037.
\bibitem{Mirzabekian:1995qf} 
  A.~G.~Mirzabekian, G.~A.~Vilkovisky and V.~V.~Zhytnikov,
  %``Partial summation of the nonlocal expansion for the gravitational effective action in four-dimensions,''
  Phys.\ Lett.\ {\bf B 369} (1996) 215,
 % doi:10.1016/0370-2693(95)01527-2
  [hep-th/9510205].
\bibitem{Mott} P.O.~Mazur and E.~Mottola, Phys.Rev. {\bf D64}(2001)104022
\bibitem{BV} A.O.~Barvinsky and G.A.~Vilkovisky, Nucl. Phys. {\bf B282} (1987) 163;
            Nucl. Phys. {\bf B333}(1990) 471)
\bibitem{FradkinTseytlin} 
  E.~S.~Fradkin and A.~A.~Tseytlin,
  %``Asymptotic Freedom In Extended Conformal Supergravities,''
  Phys.\ Lett.\  {\bf 110B} (1982) 117.
\bibitem{Paneitz} 
 S.~Paneitz ,
  %``A Quartic Conformally Covariant Differential operator for arbitrary pseudo-Riemannian manifolds,''
  MIT Preprint (1983).
\bibitem{OP} H.~Osborn and A.C.~Petkou, Annals Phys. {\bf 231} (1994) 311 
\bibitem{Erdmenger} J.~Erdmenger, Class. Quant. Grav. {\bf 14} (1997) 2061  
\bibitem{MottReview} E.~Mottola, {\tt arXiv:1006.3567[gr-qc]}   
\bibitem{Mottola06} 
  E.~Mottola and R.~Vaulin,
  %``Macroscopic Effects of the Quantum Trace Anomaly,''
  Phys.\ Rev.\ D {\bf 74}, 064004 (2006)
 % doi:10.1103/PhysRevD.74.064004
  [gr-qc/0604051].
\bibitem{Giannotti08} 
  M.~Giannotti and E.~Mottola,
  %``The Trace Anomaly and Massless Scalar Degrees of Freedom in Gravity,''
  Phys.\ Rev.\ D {\bf 79}, 045014 (2009)
 % doi:10.1103/PhysRevD.79.045014
  [arXiv:0812.0351 [hep-th]].  
\bibitem{Mott1} E.~Mottola, {\tt arXiv:1606.09220} 
\bibitem{P} A.M.~Polyakov, Phys. Lett. {\bf B103} (1981) 207
\bibitem{Tr} A. Trautman A, Bull. Acad. Polon. Sci. {\bf 6} (1958) 407 
%(reprinted as arXiv:1604.03145)
\bibitem{Waylen} P.C.~Waylen, Proc.R.Soc.London, {\bf A362} (1978) 233   
\bibitem{Deser} 
  S.~Deser,
  %``Conformal anomalies: Recent progress,''
  Helv.\ Phys.\ Acta {\bf 69}, no. 4, 570 (1996)
  [hep-th/9609138].
  %%CITATION = HEP-TH/9609138;%%
  %31 citations counted in INSPIRE as of 10 Nov 2016
\bibitem{Barvinsky} 
  A.~O.~Barvinsky,
  %``Serendipitous discoveries in nonlocal gravity theory,''
  Phys.\ Rev.\ D {\bf 85}, 104018 (2012)
 % doi:10.1103/PhysRevD.85.104018
  [arXiv:1112.4340 [hep-th]].
\bibitem{SoussaWoodard} 
  M.~E.~Soussa and R.~P.~Woodard,
  %``A Nonlocal metric formulation of MOND,''
  Class.\ Quant.\ Grav.\  {\bf 20}, 2737 (2003)
  %doi:10.1088/0264-9381/20/13/321
  [astro-ph/0302030].  
\end{thebibliography}
\end{document}